# Refugee Resettlement Housing Scout

Unaiza Ahsan
Georgia Institute of Technology
Atlanta, GA, USA
uahsan3@gatech.edu

Oleksandra Sopova
Kansas State University
Manhattan, KS, USA
osopova@ksu.edu

Wes Stayton
Georgia Institute of Technology
Atlanta, GA, USA
jstayton3@gatech.edu

Bistra Dilkina
Georgia Institute of Technology
Atlanta, GA, USA
bdilkina@cc.gatech.edu

## ABSTRACT
According to the United States High Commission for Refugees (UNHCr), there are 65.3 million forcibly displaced people in the world today, 21.5 million of them being refugees. This has led to an unprecedented refugee crisis which has led countries to accept refugee families and to resettle them. Diverse agencies are helping refugees coming to US to resettle and start their new life in the country. One of the first and most challenging steps of this process is to find affordable housing that also meets a suite of additional constraints and priorities. These include being within a mile of public transportation and near schools, faith centers and international grocery stores. We detail an interactive data-driven web-based tool, which incorporates in one consolidated platform most of the needed information. The tool searches, filters and demonstrates a list of possible housing locations, and allows for the dynamic prioritization based on user-specified importance weights on the diverse criteria. The platform was created in a partnership with New American Pathways, a nonprofit that supports refugee resettlement in the metro Atlanta, but exemplifies a methodology that can help many other organizations with similar goals.

## 1. INTRODUCTION
There are more than 65 million forcibly displaced people in the world today – more than at any other time in recorded history [1]. People are forced to seek a safer place to live and raise their families because of frequent wars, constant political and economic crisis and possible persecution. Essentially, these people become refugees and some of them are resettled in richer countries of Europe, Asia and the US. Up to 70,000 refugees are resettled in the US with approximately 2,500 going to Georgia annually.

We have been working closely with the New American Pathways non-profit organization that helps refugees coming to Atlanta to resettle and start their new life. One of the first and most challenging steps of this process is to find affordable housing that also meets a suite of additional constraints and priorities such as proximity to public transit (since most refugees do not own a vehicle), schools (since most of them have children of school age), supermarkets (halal, international), ESL classes, faith communities and other. Also, another major constraint is that refugees often do not the documents e.g. SSN necessary to access some special types of housing.

Taking all these constraints into account, finding safe yet affordable housing for refugees becomes non-trivial. Currently, the process of finding suitable housing for incoming refugees is a manual, tedious task. The resettlement agency begins the search for apartments and other relevant information on the Internet, and then working step by step to check if most important constraints are satisfied. For example, the crime situation near an apartment is validated by physically visiting the area and by word-of-mouth reports. In this paper, we present our solution to optimize the process by developing an interactive data-driven web-based tool, which searches, filters and presents a list of optimal housing locations, and allows for the dynamic prioritization of apartment locations based on user-specified importance weights on the diverse selection criteria. The tool is intended to be used by refugee resettlement agencies in their search to find appropriate apartments for refugee families just arriving into the country. The agencies can also use this tool to help refugee families in search for apartments when they decide to move from their intake apartment placement, usually within the first or second year after arrival.

## 2. APPROACH
We approach the problem of finding optimal housing by first assessing the data requirements. There are specific constraints refugee resettlement agencies face when they arrive in the US. The incoming families are on a special visa, do not have social security cards, have little to no resources of their own, do not own a vehicle and are generally dependent on the resettlement agencies they are associated with. This means that the two of the highest priorities for any resettlement location are the rent and access to public transportation. Next in line is access to schools and then international grocery stores, hospitals, faith communities, English as Second Language (ESL) classes, Department of Driver Services (DDS) offices, Social Security Number (SSN) offices etc.

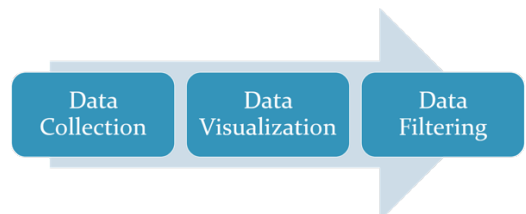

**Figure 1: Approach taken in this paper**

After data collection, we use mapping tools to visualize the data and finally develop an interactive web application that is able to filter out irrelevant pieces of information and display optimal locations for refugee resettlement.

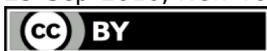

Table 1: Sample Data Collected Using Google Places API

| Place Name | Place Type | latitude | longitude | Place Address | Phone | Zipcode |
|---|---|---|---|---|---|---|
| 10 Perimeter Park Apartments | Real estate | 33.918 | -84.294 | 10 Perimeter Park Dr, Atlanta, GA | (770) 458-1001 | 30341 |
| Zamzam International Foods | Grocery/supermarket | 33.792 | -84.229 | 5030 Memorial Dr, Stone Mountain, GA | (404) 294-0911 | 30083 |
| New Hope Primary Care | health | 33.934 | -83.956 | 930 New Hope Rd #10, Lawrenceville, GA | (770) 868-5160 | 30045 |
| Temple Sinai | synagogue | 33.910 | -84.417 | 5645 Dupree Dr NW, Sandy Springs, GA | (404) 349-8956 | 30327 |
| ESL Adult Education Class | ESL | 33.524 | -84.358 | 137 Spring St, Jonesboro, GA | (770) 515-7610 | 30236 |
| Briarcliff Child Care Center | daycare | 33.829 | -84.33 | 2260 N Druid Hills Rd NE, Atlanta, GA | (404) 636-3504 | 30329 |

## 3. DATA COLLECTION

Given the resettlement criteria and related information, the task of data collection can be divided into two parts:

1. Places data: This includes all apartment complexes, schools, grocery stores, hospitals, faith centers and their details. We use Google Places API to collect this data.
2. Area Attribute data: This includes crime information, affordability of housing by area, job and retail access.

### 3.1. PLACES DATA COLLECTION

We use the Google Places API's text search feature to generate places and their details for a particular area. The text search method takes as input the API key (which is obtained freely from Google Places API), the location of search and search text query. Our location of search are two counties in Georgia: Fulton and DeKalb counties. Hence we obtain a list of zip codes that belong to these two counties. Then we input the text search query term "apartment complexes in <zip code>" along with location "Georgia, United States" and the API key in the call to the API for place search. This API call results in a search result which gives basic information about a place namely, its name, latitude, longitude and a unique place ID. The place ID is used to make another API call (place.get_details()) and this is used to get further details about each place. Further details include: place address, phone number, website URL, opening hours (if it is a commercial address), etc.

Table 1 shows sample data collected using Google Places API. Note that schools are not included in this data. The schools and their details were collected using Great Schools API. The API is rich in that it provides not only the basic information about a school, but also its reviews, its ethnicities and the percentage of students who use the free and reduced lunch program. This information is quite important for our application. These details are displayed in a popup as a user clicks on a school icon on the map.

### 3.2. AREA ATTRIBUTE DATA COLLECTION

Block group level data on affordability, retail access, and job access was provided by the 2014 Location Affordability study done by the Department of Housing and Urban Development [2]. Job and Retail access were taken from their respective indices in the data set while the affordability estimate came from a combination of variables. To approximate the housing costs we used the estimated percentage of income spent on housing for a two parent, two child household and multiplied that by the median annual income for that block group. We then take the annual housing costs and divide that by 12 to get a monthly estimate. We then pre-filtered out all block groups with average costs above $3000 per month because they are not viable options.

Although the New American Pathways organization has a price ceiling of roughly $1000 per month we opted to include block groups with substantially higher prices so as not to eliminate any possible affordable enclaves within more expensive block groups. The apartments that we pulled have a median monthly cost of $1147 and a standard deviation of $582.

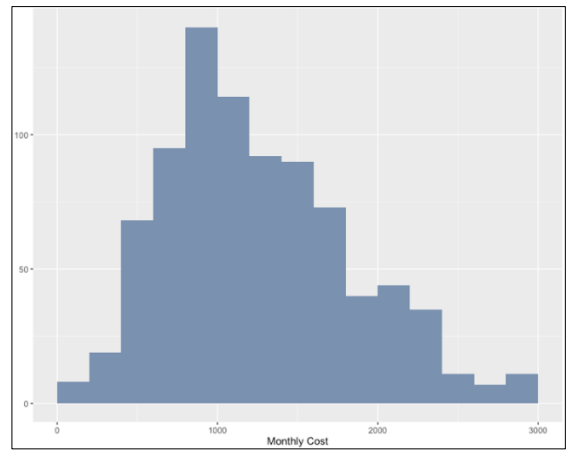

Figure 2: Distribution Showing Monthly Costs of Apartments

Initially, we explored pulling affordability numbers and actual live apartment listings from Zillow and Trulia but through client feedback and our own research this proved problematic for two reasons. First, listings are not up to date on either website, that is, most listings are just placeholder listings for complexes which often have units available but not always, and not at the specified price range. Second, prices are often inaccurate and understate the actual rents. Our investigation showed that prices were shown at discounts of 10 - 50 % from actual prices given by the complexes themselves.

For this reason, we chose to simply include all apartments in DeKalb and Fulton counties and let the end users do their own investigation with the help of our cost estimation.

## 4. DATA VISUALIZATION

Our interactive tool consists of the following elements:

We developed a web-based interactive tool for visualization of all of the data that we have collected and cleaned, which include:

- Apartment complexes in Fulton and DeKalb counties, (including a name, a link to the website, a phone number, an address and other contact information, and travel time to the New American Pathways (NAP) office, since refugees need to visit the NAP office frequently during the first six month)

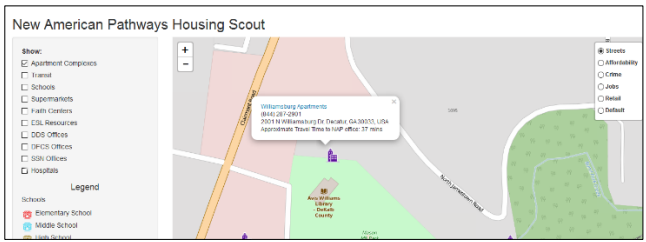

**Figure 3: Visualization of Apartment Complex Details**

- Public transit (MARTA) stops

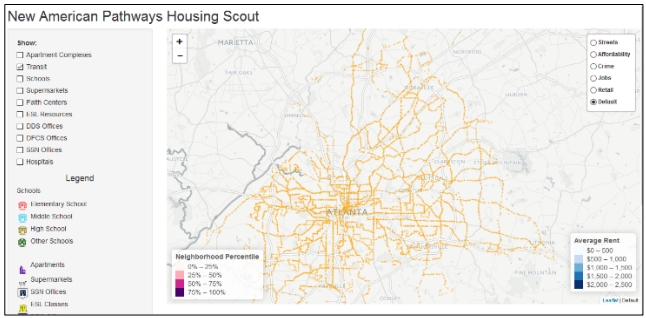

**Figure 4: Visualization of all transit stops in Fulton and DeKalb Counties, GA**

- Schools (including public/private status, percentage of free/reduced price lunch, a phone number, an address and other contact information).

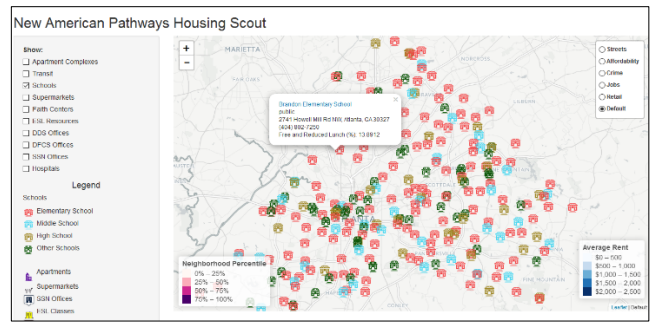

**Figure 5: Visualization of Schools in Fulton and DeKalb Counties**

- Supermarkets (including a name and an address and other contact information)
- Faith centers (including a name and an address, also differentiation between Christian churches, Muslim mosques, Hindu temples and other)

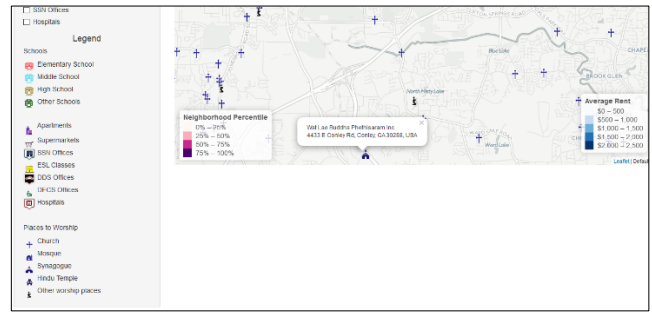

**Figure 6: Visualization of Faith Centers in Fulton and DeKalb Counties**

- English as Second Language (ESL) resources (including a name, an address and other contact information)
- Department of Driver Services (DDS) offices
- Division of Family and Children Services (DFCS) offices
- Social Security (SSN) offices (including a name, a phone number, an address and other contact information)
- Hospitals (including a name, working hours, a phone number, an address and other contact information)

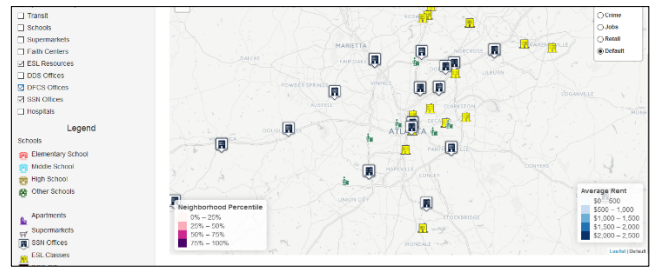

**Figure 7: Visualization of ESL Resources, DFCS offices and SSN offices**

In addition, we create six different layers ("Streets", "Affordability", "Crime", "Jobs", "Retail", "Default") that give relevant information (expressed in percentiles) about the areas in Fulton and DeKalb counties.

We also provide an opportunity for the user to prioritize different aspects of apartment complexes' locations, such as proximity to supermarkets, public transit, and schools.

On startup, the user is presented with a map of Atlanta with tiles provided by OpenStreetMap. They can then add any of the data located on the selection panel to the left of the map. They can also change their base map (on the right; see Figure 6) to show one of the layers we have provided, e.g. "Crime", "Retail" etc.

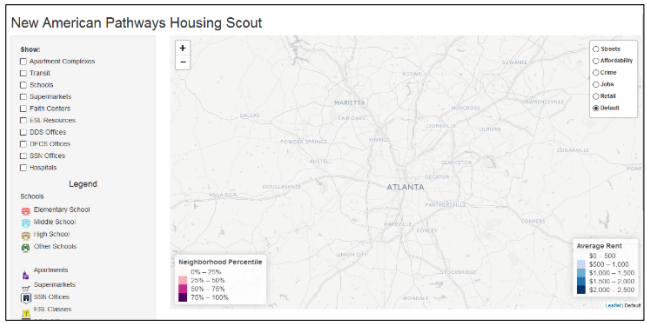

**Figure 8: Landing Page**

We recommend using the default map to get a lay of the land and then proceeding to the "Affordability" to quickly identify areas which conform to our primary constraint, price. After getting a sense of the most suitable areas we can go switch to the "Retail" and "Jobs" layers to further narrow our search. It is worth noting that the "Retail" and "Jobs" layers (legends to the bottom right; see Figure 8) compare areas of Atlanta against one another, providing a relative ranking of suitability.

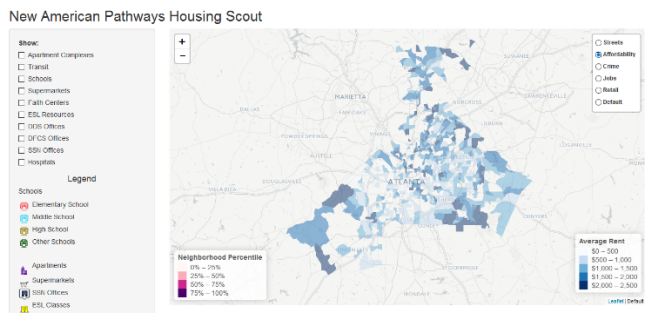

**Figure 9: Visualizing the "Affordability" Layer**

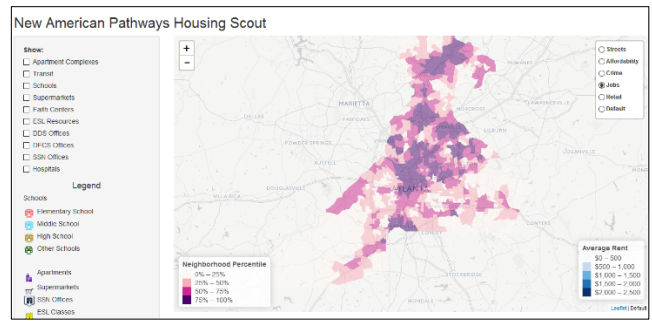

**Figure 10: Visualizing the "Jobs" Layer**

After identifying a promising area, we can then investigate the apartment complexes in the area. At this point we will want to add our secondary selection criteria - "Transit", "Schools", and "Markets" layers - to the map. After finding a suitable apartment, we can then click on the apartment icon to bring up additional details, most importantly, the web link to the apartment complex website where the user can do their own research and obtain actual price information.

## 5.DATA FILTERING

The user can also use the sliders to specify their own priorities. Sliders allow the user to assign weights to the attributes of the areas on the map, including Affordability, Jobs, Retail, and proximity to public transit, the New American Pathways HQ, and Schools. After assigning weights the user will be presented with a custom raster heat map showing the desirability of each area of the map. Additionally, they will be shown an ordered list of complexes which most suit their selection preferences.

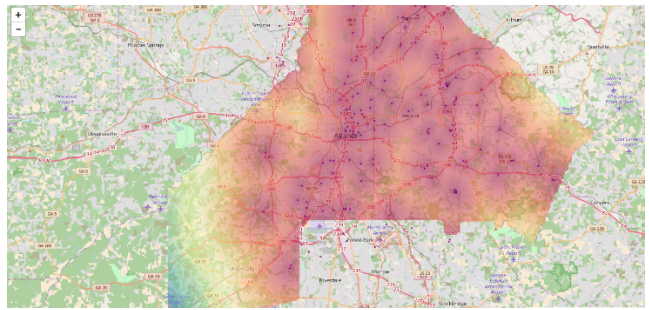

**Figure 11: A Heat Map Emphasizing Proximity to Markets**

## 6.UPDATING APARTMENT LISTINGS

We provide the option of updating the apartment listing to the user via an update form. The update form takes as input the name of the apartment and other details such as its address, phone number and website.

**Figure 12: Form to Add New Apartments to the Map**

This information is then stored in a CSV file and finally added to the main apartment listings file. To prevent overwriting information if multiple users submit the form, we do two things. We choose the submission filename to be a concatenation of the current time and the md5 hash of the submission data. Hence he only way two submissions will overwrite each other is if they have the exact same contents and happen at the exact same second. Figure 11 shows the form used for letting users add new apartments to the listing. Note that the apartment name and address are mandatory fields. This is because we need the address of a place in order to geocode it and display on the map and the apartment's name is its identifying information.

## 7.CONCLUSION

Our project combines interactive visualizations with one stop shop information gathering to provide users with the information they need to optimize both area and apartment selection. Once deployed, the housing scout will be a tool which can lighten the manual workload of housing coordinators and provide intelligence for more finding ideally suited resettlement locations. The tool is limited to the Atlanta area for now but provides a blueprint for the development of tools to help refugee resettlement agencies in other parts of the country.

## 8.ACKNOWLEDGEMENTS

We acknowledge the Data Science for Social Good Atlanta summer program, which supported this project and the three students involved. We gratefully acknowledge Alex Godwin and Prof. Polo Chau for their valuable input.

## 9.REFERENCES

[1] http://www.unhcr.org/en-us/figures- at-a- glance.html

[2] Department of Housing and Urban Development, Location Affordability Portal - Version 2. (accessed June 15, 2016), www.locationaffordability.info/about.aspx